\documentclass[a4paper]{mem}
\usepackage{natbib}
\usepackage{graphicx}
\usepackage[a4paper]{hyperref}
\idline{73}{23}
\begin{document}
\title{Field and cluster RR Lyrae stars as stellar tracers}
%\thanks{this is a place for a title footnote}
%}

\author{S. Petroni \inst{1} 
\and G. Bono \inst{2}}
          %\fnmsep
%\thanks{this is a place for placing a footnote in the author field }
%}

%%   \offprints{S. Petroni} \mail{Dipartimento di Fisica, via Buonarroti
%%2, Pisa, 56127 Italy}

\institute{Dipartimento di Fisica, Universit\`a degli Studi di
Pisa, via Buonarroti 2, Pisa, 56127 Italy;
\email{petroni@df.unipi.it}\\ \and INAF - Osservatorio Astronomico di
   Roma, via Frascati 33, 00040 Monte Porzio Catone, Italy; \email{bono@coma.mporzio.astro.it}}

\abstract{We present an overview on pulsation properties of RR Lyrae
stars in the Galaxy and in Local Group dwarf galaxies. We discuss the 
key information that RR Lyrae might provide on their parent population, 
and in particular on their metallicity distribution. 
  
   }
   \authorrunning{S. Petroni and G. Bono}
   \titlerunning{Field and cluster RR Lyrae stars as stellar tracers}
%%   \maketitle
%
%________________________________________________________________

\section*{{\Large Field and cluster RR Lyrae stars as stellar tracers}}

\vspace{0.7cm}
{\bf S. Petroni}

\noindent
{\small Dipartimento di Fisica, Universit\`a degli Studi di Pisa, via
Buonarroti 2, Pisa, 56127 Italy; email: petroni@df.unipi.it}\\

\noindent
{\bf G. Bono}
 
\noindent
{\small INAF - Osservatorio Astronomico di Roma, via Frascati 33,
00040 Monte Porzio Catone, Italy; email: bono@coma.mporzio.astro.it}\\

\section*{Abstract}

We present an overview on pulsation properties of RR Lyrae
stars in the Galaxy and in Local Group dwarf galaxies. We discuss the 
key information that RR Lyrae might provide on their parent population, 
and in particular on their metallicity distribution.

\section{Introduction}

The large photometric databases collected by microlensing projects
(EROS, MACHO, OGLE, PLANET) substantially increased the number of
variable stars for which are available accurate estimates of pulsation
properties. This provided the unique opportunity to investigate in
detail the occurrence of peculiar phenomena such as Bump and Beat
Cepheids, Blazhko RR Lyrae, and mixed-mode variables.  Even though
their existence was predicted long time ago, second overtone Cepheids
have also been recently discovered in the Small Magellanic Cloud (SMC)
\citep{uall99}. On the basis of a sample of 7900 field RR Lyrae stars
observed in the bar of the Large Magellanic Cloud (LMC),
\citet{alcock96} also suggested the identification of second overtone
pulsators.  This finding relies on the evidence that the period
distribution of first overtone RR Lyrae (RRc) presents a secondary
peak located at $P\approx 0.281$ days. In the following we will call
this peak as the {\em Macho peak}.

On the other hand, current nonlinear, convective models \citep{bono97}
suggest that such a peak might be the signature of a more metal-rich
population of RR Lyrae stars.  A similar feature in the period
distribution of RRc stars was also detected by \citet{kal95} in the
Sculptor dwarf spheroidal (dSph) galaxy. The {\em macho peak} in this
sample was located at $P\approx 0.290$ days. According to theoretical
and empirical evidence, fundamental RR Lyrae (RRab) present an
anti-correlation between period and amplitude, i.e. RRab stars with
longer periods present smaller luminosity amplitudes. On the other
hand, RRc stars present, in the Bailey diagram (luminosity amplitude
vs period), a distribution that resembles a ``bell'' shape
\citep{bono99}. The Bailey diagram of RRc variables in Sculptor shows
two shifted ``bell'' shapes, and the bulk of stars are located across
the two peaks detected in the period distribution.

Theoretical and empirical findings also bring forward that the
distribution of RRab stars in the Bailey diagram is marginally
affected by metal content for $[Fe/H]\le-0.6$. Oddly enough, current
theoretical predictions do suggest that an increase in the metal
abundance causes, at fixed input parameters, a decrease in the mean
luminosity and, in turn, a decrease in the pulsation period of RRc
stars.  The aftermath of this change is that two RRc samples
characterized by two different mean metallicities disclose two
different ``bell'' shapes in the Bailey diagram. Therefore the
occurrence of a broad distribution in the Bailey diagram and in the
period distribution of RRc stars might be the fingerprint that the
underlying stellar population presents a large spread in
metallicity. RR Lyrae variables in Sculptor appear as the template for
this effect, since current spectroscopic measurements support the
evidence that its mean metal content is $\langle [Fe/H] \rangle
\simeq -1.5$ dex with a dispersion of $\sim \pm 0.9$ dex
\citep{tall02}.  The same outcome applies for RR Lyrae in LMC, since
spectroscopic measurements of 15 field RR Lyrae stars indicate a
$\langle [Fe/H] \rangle = -1.7$ with tails extending to -0.8 and to
-2.4 \citep{alcock96}.

To investigate in more detail whether this finding is an intrinsic
feature of RR Lyrae in these two stellar systems, we decided to
investigate the pulsation properties of RR Lyrae in Local Group (LG)
dwarf galaxies, since they present a broad dispersion in chemical
composition.

%%%%%%%%%%%%%%%%%%%%%%%%%%%%%%%%%%%%%%%%%%%%%%%%%%%%%%%%%%%%%%%%%%%%%%%%%%%%%%
\section{Galactic and Local Group stellar systems}

A glance at the period distribution of RR Lyrae in two populated
Galactic globular clusters (GGCs), M3 and M15, i.e. the prototypes of
Oosterhoff type I (Oo I) and Oosterhoff type II (Oo II)
groups\footnote{According to Oosterhoff (1939, 1944) GGCs that host 
RRab variables with a mean period $\approx 0.55$ days are called type I, 
while those with a mean period $\approx 0.65$ days are called type II 
clusters.}, discloses that no firm conclusion can be drawn due to the 
limited number of RRc variables. The only exception to this rule is
$\omega$~Cen. This is a massive globular cluster (total luminosity
$M_V=-10.29$ mag; \citealt{h96}) and presents a well-defined
metallicity spread (\citealt{rall00}; \citealt{pall02}). Even though
the census of RR Lyrae stars in this cluster is far from being
complete, the period distribution of RR Lyrae stars collected by
\citet{kal97} seems to show the {\em Macho peak}.  The problem
of limited samples in individual GGCs might be overcome by taking into
account the cumulative period distribution of RR Lyrae stars in GGCs
\citep{cle01}. Both Oo I and Oo II GGCs show the {\em Macho peak} (see
their Fig. 3). Interestingly enough, this secondary feature is more
evident among Oo II clusters which present a larger spread in metal
abundance, namely $-2.29 \le [Fe/H] \le -0.28$.

Now, we move our analysis to LG dwarf galaxies. The reasons are manifold: 
{\em i)} a large fraction of them hosts sizable samples of RR Lyrae stars 
(Mateo 1998); {\em ii)} the dynamical and star formation history of these 
stellar systems are substantially different than GGCs, and therefore they 
supply the unique opportunity to investigate whether these properties 
affects the pulsation properties of RR Lyrae; {\em iii)} recent results 
based on the radial distribution of RR Lyrae stars in the Galactic halo 
(\citealt{vivas01}, see also the paper by Vivas at these proceedings) 
strongly support the evidence that the clump of 
stars detected by the Sloan Digital Sky Survey (\citealt{sdss2}), at 
approximately 50 kpc from the Galactic center, might be the tidal 
stream left over by the Sagittarius (Sgr) dSph. 

This means that pulsation properties of RR Lyrae stars in the halo
clump can be soundly adopted to single out whether they do belong to
the Galactic halo or to Sagittarius. Data plotted in Fig. 1 clearly
show that the period distribution of RRc in Sagittarius presents a
well-defined peak centered on $P\approx0.32$ days. On the other hand,
data available in the literature indicate that RRc in the halo present a
broad period distribution ranging from 0.2 to 0.4 days and do not show
a well defined peak. Oddly enough, the spread in metallicity of RRc
stars in the halo is narrower ($\langle [Fe/H] \rangle \simeq -1.65$
dex with a dispersion of $\sim \pm 0.3$ dex; \citealt{skk91}) than in
Sagittarius ($\langle [Fe/H] \rangle \simeq -1.0$ dex with a
dispersion of $\sim \pm 0.8$ dex; \citealt{tall02}). 
Unfortunately, we still lack detailed information concerning the
period distribution and the luminosity amplitude of RRc stars in the
new clump, and therefore we cannot perform a detailed comparison with
quoted samples.
Moreover, both the mean and the spread in chemical composition in 
dwarf spheroidals are widely debated, since the difference between 
spectroscopic measurements and photometric indexes is larger than 
current uncertainties \citep{tall01}. 
At present, it is not clear whether such a discrepancy is caused by a
different evolution either of Iron-peak elements or of
$\alpha$-elements such as Calcium. Finally, we note that the large
spread in CaII triplet metal abundances might also be due to a
systematic difference among intrinsic stellar parameters such as
gravity and temperature \citep{eb91}.
  
%
%                                                Two column figure
%----------------------------------------------------------- S_vib   
\begin{figure*}
\centering
\resizebox{\hsize}{!}{\includegraphics[]{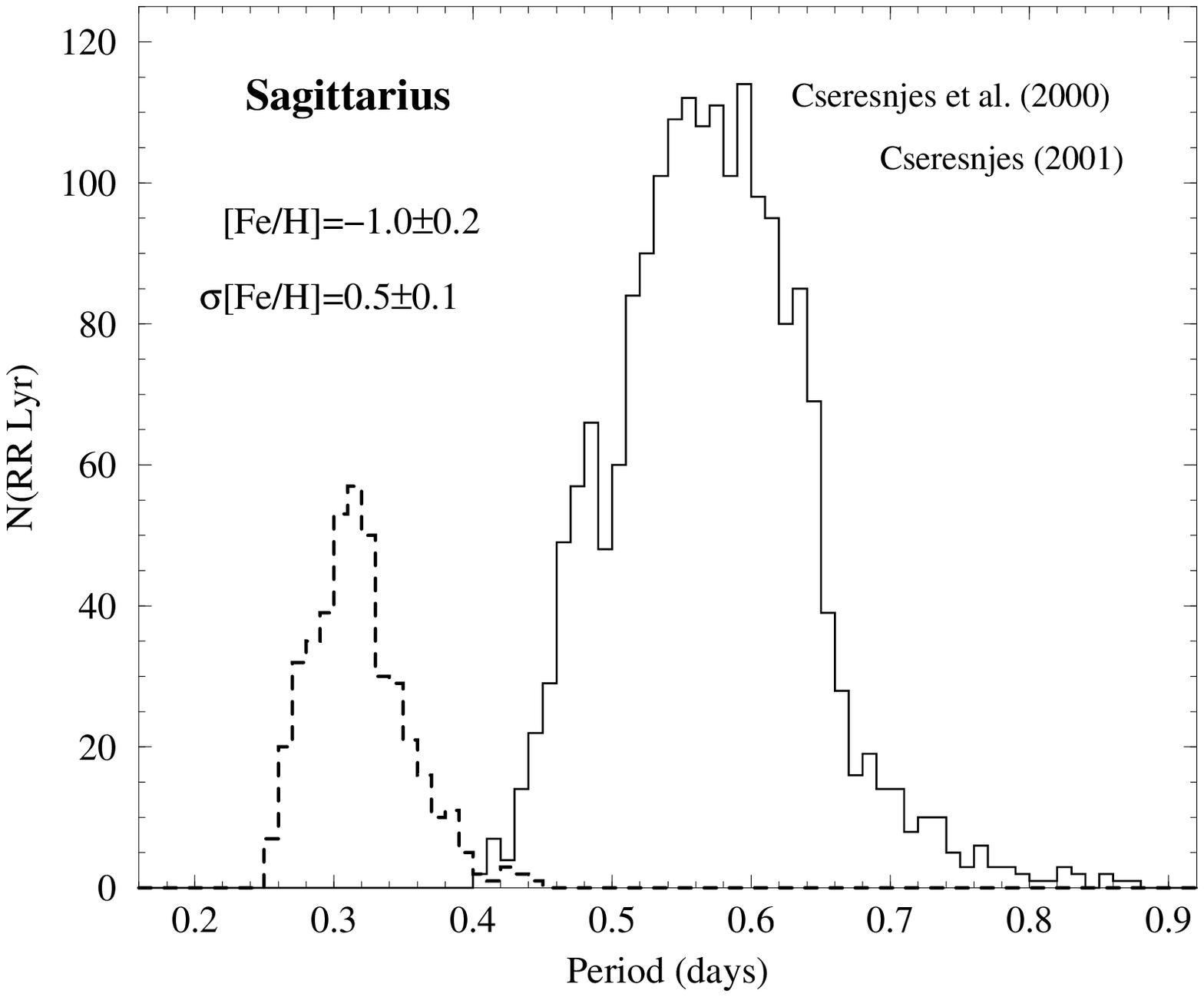}
\includegraphics[]{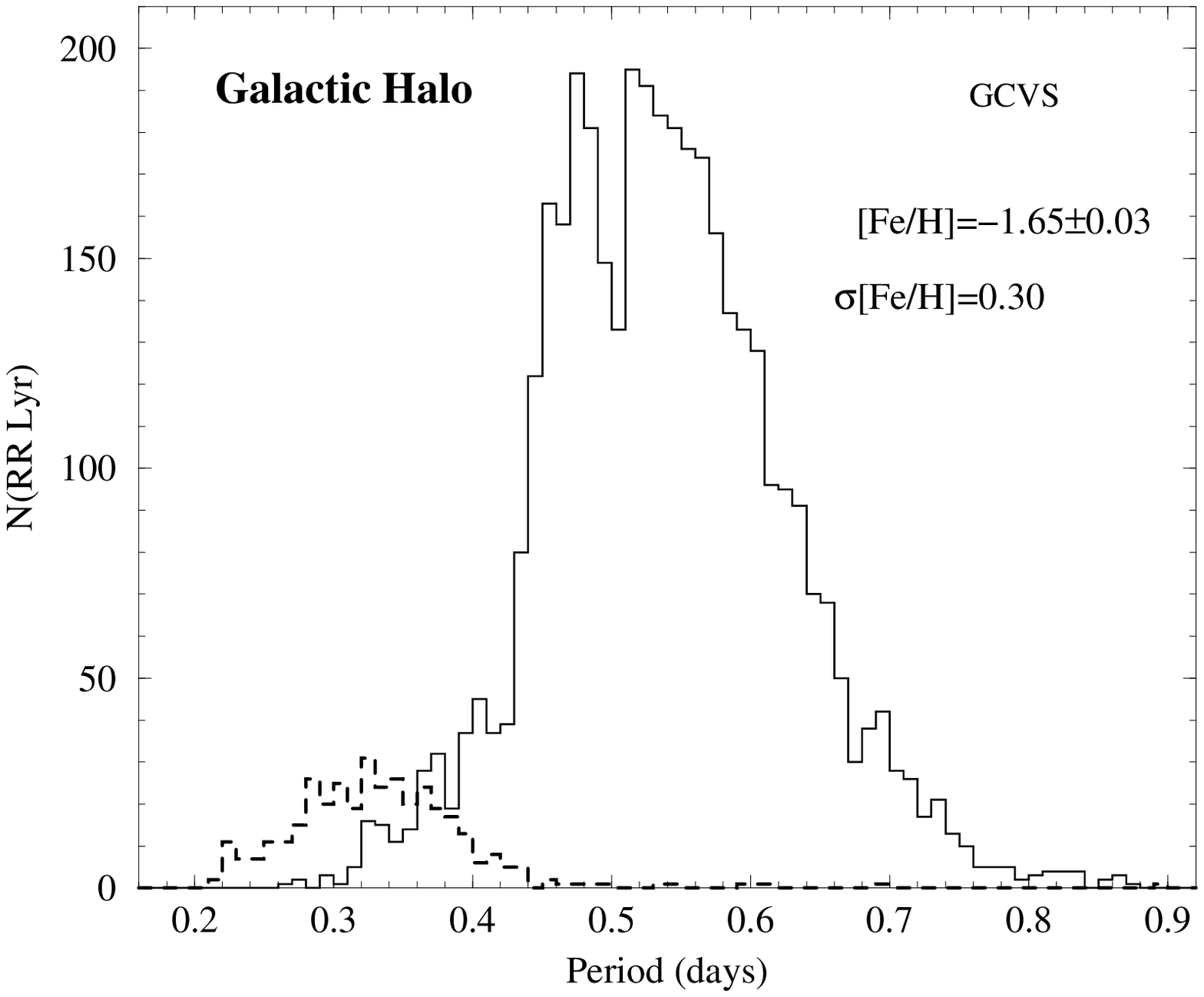}}
\caption{Period distribution of RR Lyrae stars in 
the Sgr dSph \citep{patrik00,patrik01} ({\it left}) and in the Galactic halo, 
(General Catalogue of Variable Stars, GCVS; {\it right}).}
   \label{f1} \end{figure*}
%
%______________________________________________________________
%
This not withstanding, preliminary results by \citet{vivas01} suggest
a mean period of RRab stars in the clump of $\langle P\rangle=0.58$
days. Interestingly enough, this value is very much similar to the
mean period of RRab in Sgr, $\langle P\rangle=0.57$ days
\citep{patrik01}, than to RRab in the Galactic halo, i.e. $\langle
P\rangle=0.54$ days (General Catalogue of Variable Stars, GCVS). This
finding is further strengthened by the fact that recent data collected
by \citet{patrik01} for RR Lyrae stars toward Sagittarius (Galactic
Center, namely a mix of thick disk and bulge stars), present a mean
period of RRab variables similar to the Galactic halo, i.e. $\langle
P\rangle=0.55$ days.

%
%                                                Two column figure
%----------------------------------------------------------- S_vib   
\begin{figure*}
\centering
\resizebox{\hsize}{!}{\includegraphics[]{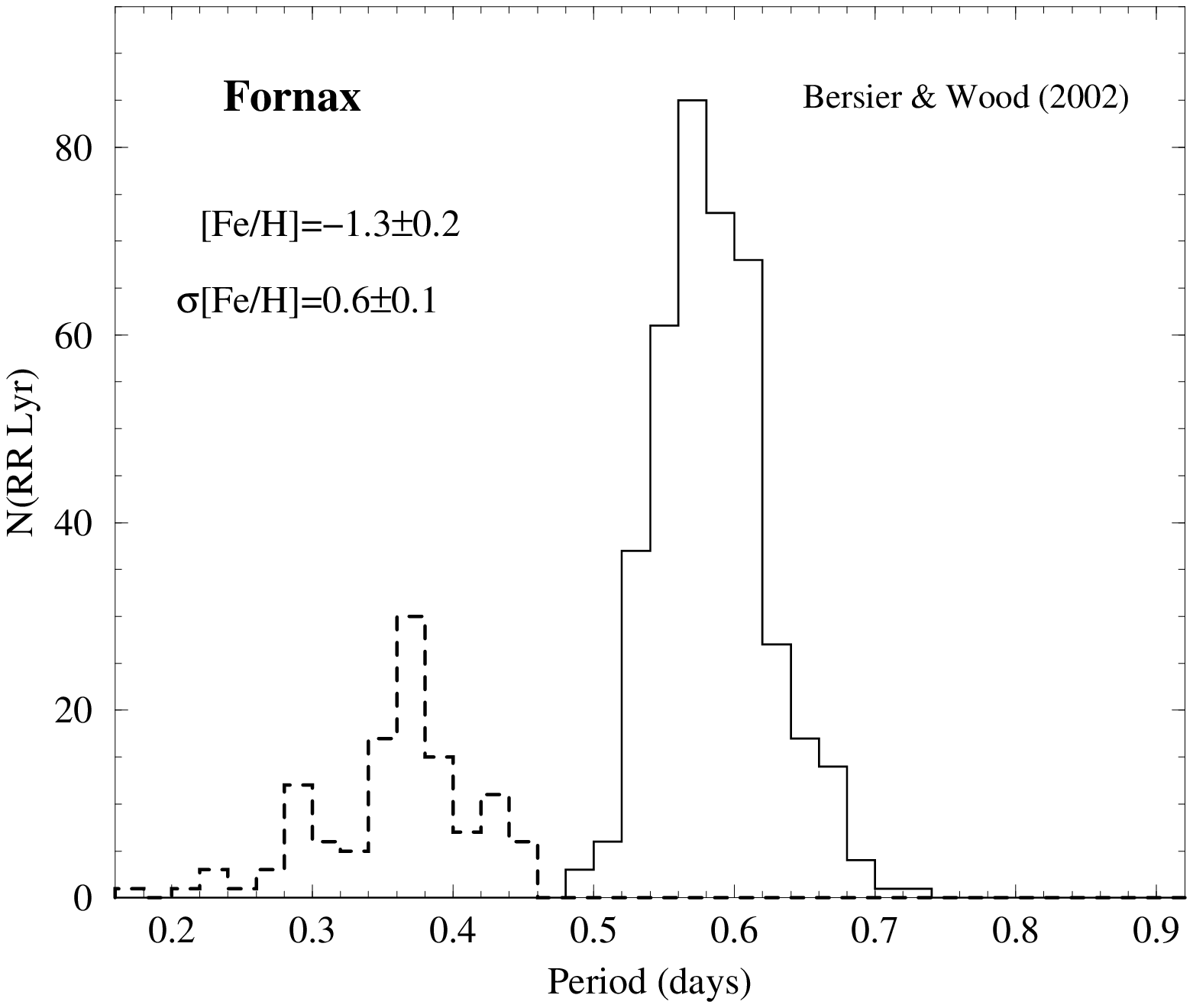}
\includegraphics[]{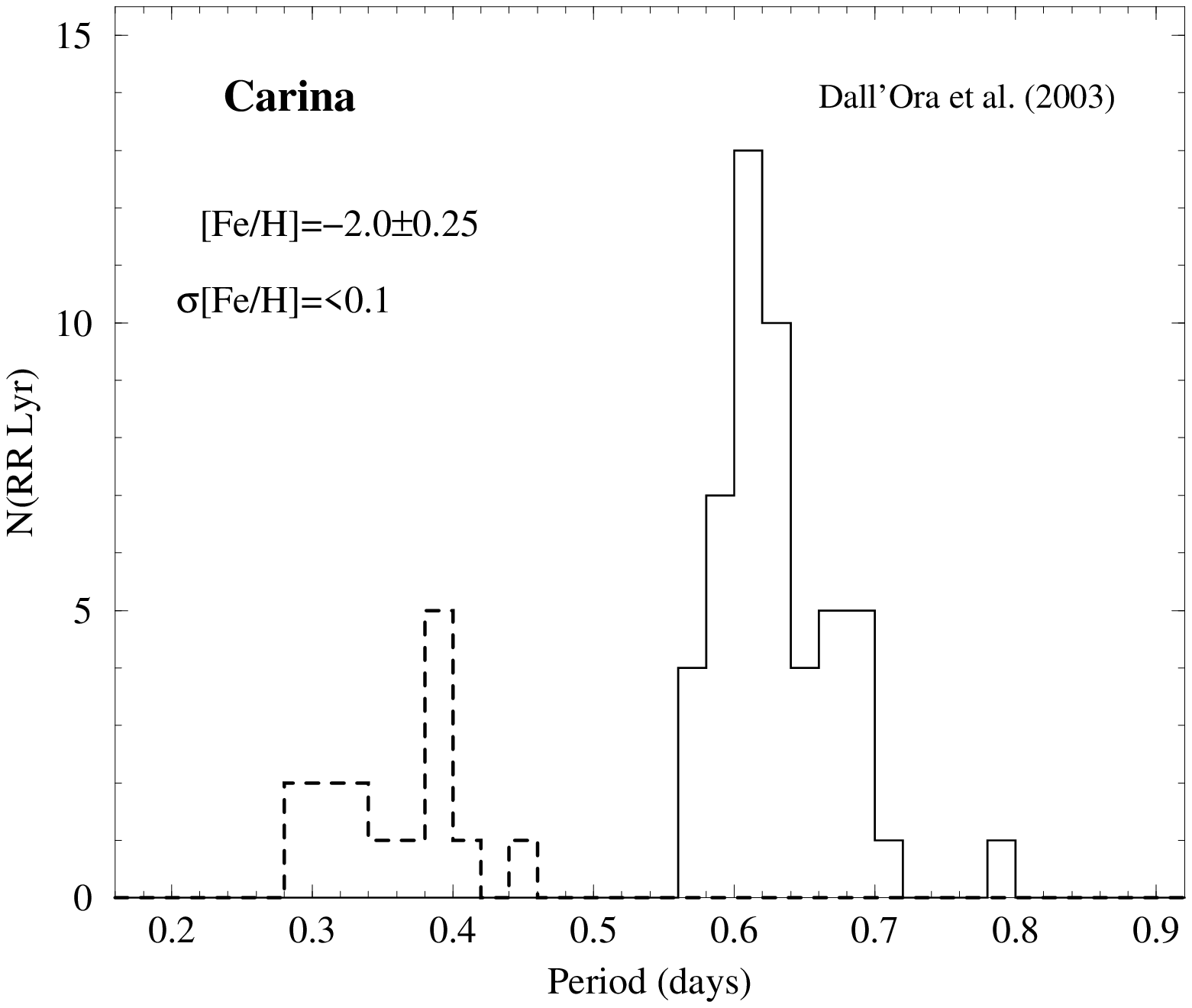}}
\caption{Period distribution of RR Lyrae in Fornax 
\citep{fornax} ({\it left}) and in Carina \citep{carina} ({\it right}) 
dSph galaxies.}  \label{f2}
\end{figure*}
%
%______________________________________________________________
%
Data plotted in Fig. 2, together with data collected by
\citet{patrik01} (see his Figure~5) 
suggest a tight connection between the occurrence of the {\em Macho peak}
in the period distribution of RRc stars and a large spread in metal content. 
The only exception to this rule seems to be the Sgr dSph. Whether this 
evidence could cast some doubts on the observed spread in metallicity, it 
is not clear yet. Note that for several LG dwarfs the sample of RRc 
stars is too small (Carina) or the entire sample is not complete (Draco, 
Sextans).  

\begin{figure*}
   \centering
\resizebox{\hsize}{!}{\includegraphics{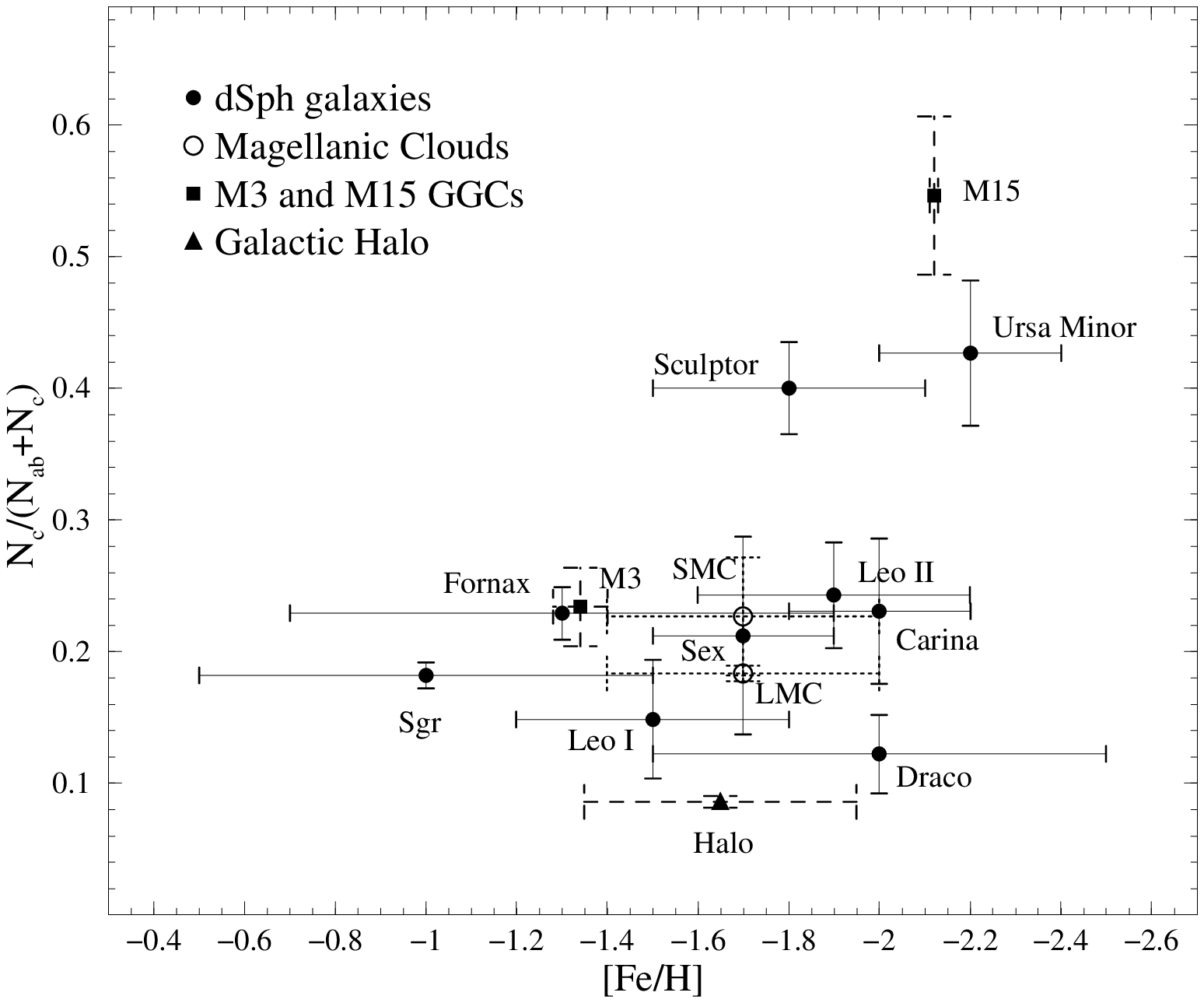}}
\caption{Relative number of first-overtones (RRc) to the total number 
of RR Lyrae for the labeled stellar systems. Data were collected by: 
\citet{carina} (Carina); \citet{n85} (Draco); \citet{h01} (Leo I); 
\citet{sm00} (Leo II); \citet{kal95} (Sculptor); 
\citet{mfk95} (Sextans); \citet{nwo88} (Ursa Minor); 
\citet{patrik01} (Sagittarius); \citet{fornax} (Fornax); 
\citet{alcock96} (LMC); \citet{g75}, \citet{sall92} (SMC); 
GCVS (halo); \citet{cc01} (M3); \citet{ss95} (M15). Metallicity
spreads are from \citet{m98}.}\label{f3}%
    \end{figure*}

To investigate in more detail the problem we decided to adopt a new 
observable, namely the ratio between RRc stars and the total number of 
RR Lyrae. Figure~\ref{f3} shows this ratio as a function of the metal 
content for stellar systems in the Galaxy and in the LG. 

%    {!}                                 Two column figure (place early!)
%______________________________________________ Gamma_1 (lg rho, lg e)
%______________________________________________________________
Data plotted in this figure show a substantial difference between 
the relative number of RRc stars in Sgr and in the halo. Although, 
RRc in the halo might be affected by a selection bias, this parameter 
seems quite promising to split RR Lyrae stars belonging to different 
parent populations.

%%%%%%%%%%%%%%%%%%%%%%%%%%%%%%%%%%%%%%%%%%%%%%%%%%%%%%%%%%%%%%%%%%%%%%%%%%%%
\section{Conclusions}

The {\em Macho peak} in the period distribution of RRc variables
appears to be connected with a large spread in the metal content of
the parent population. This finding is supported by RR Lyrae in GGCs
($\omega$ Cen, Oo II clusters) and in LG dwarf galaxies. The only
exception to this empirical evidence are RR Lyrae in Sagittarius.  We
found that the mean period of RRab variables in the halo clump
detected by \citet{vivas01} agrees quite well with the mean period
of RRab in Sagittarius.  Finally, we found that the ratio between RRc
and total number of RR Lyrae seems a robust observable to figure out
whether RR Lyrae in the halo clump do belong to Sagittarius or to the
Galactic halo.

\bibliographystyle{aa}

\end{document}